# Analysis of Winske-Daughton 3D Electromagnetic Particle Simulation of Ion Ring Generated Lower Hybrid Turbulence

L. Rudakov[1], C. Crabtree, M. Mithaiwala, and G. Ganguli

Plasma Physics Division, Naval Research Laboratory

Washington D.C. 20375-5346, USA

[1]Icarus Inc., Bethesda, Icarus Research Inc., P.O. Box 30870, Bethesda, MD 20824-0780, USA

Abstract

Using electromagnetic particle-in-cell simulations Winske and Daughton [Phys Plasmas, 19, 072109, 2012] have recently demonstrated that the nonlinear evolution of a wave turbulence initiated by cold ion ring beam is vastly different in three dimensions than in two dimensions. We further analyze the Winske-Daughton three dimensional simulation data and show that the nonlinear induced scattering by thermal plasma particles is crucial for understanding the evolution of lower hybrid/whistler wave turbulence as described in the simulation.

1. Introduction

A realistic numerical simulation of the nonlinear dynamics of the evolution of an ion ring beam localized over a finite region of space continues to be challenging. The complexity arises due to nonlinear interactions of waves and particles in turbulence that is inherently three-dimensional (3D) in nature and involves nonlinear (NL) scattering of short wavelength electrostatic waves into long wavelength electromagnetic waves and vice-e-versa with large disparities in both spatial and temporal scale sizes. Realistic simulations are possible only in limited domains with assumed boundary conditions and to date no complete end-to-end simulation of this process is available. Recently Winske and Daughton[1] took a major step towards a more complete simulation of ion ring beam evolution. They presented 2D and 3D electromagnetic Particle-in-Cell (PIC) simulations of the nonlinear dynamics of the evolution of an ion ring beam in the Very Low Frequency (VLF), $\Omega_i < \omega < \Omega_e$, range where the wave frequency $\omega$ is bound by the electron and ion gyro-frequencies $\Omega_{e,i}$. Despite their heroic effort, the complete and realistic simulation of an ion ring beam formation and evolution in space starting from the injection of a jet of neutral atoms perpendicular to the ambient magnetic field, their photo-ionization into an ion ring beam, onset of quasi-electrostatic Lower Hybrid (LH) turbulence, their subsequent nonlinear scattering into electromagnetic whistler/magnetosonic waves, and the propagation of these waves away from the region of creation, etc., still eludes us. This is mainly because, for reasons of practicality, the simulation[1] had to consider an instantly ionized beam spread over the entire space as opposed to being localized in a finite extent of space and ionizing over a duration of time, as is the reality. This compromise seriously affects the applicability of the simulation[1] to the realistic problem since it does not allow the energy to propagate away from the region of creation as theoretical analysis[2] of the issue indicates. This limitation results in a distortion of the energy budget in a local region of space, which results in inaccuracies in the calculated efficiency for energy conversion. However, much has been learned from the Winske-Daughton (WD) simulations[1] while more information still remains to be gleaned from it. The purpose of this article is to further analyze the nonlinear evolution given in the WD simulations[1] in an attempt to extract more information than was originally done. In particular, we highlight the crucial role of induced NL scattering by thermal plasma particles, i.e., nonlinear Landau damping, which has not been duly recognized but is essential for an accurate understanding of the resulting signatures in the simulation.

## 2. Linear Analysis of the Simulation Results

We first discuss the linear characteristics of the waves generated in the 3D simulations[1]. We focus only on the 3D simulations since in 2D the nonlinearities are different and not realistic for the LH/whistler turbulence, which is inherently 3D in nature[2]. The dispersion relation of the VLF (i.e., whistler, lower hybrid, and magnetosonic) waves in cold plasma limit is given by[2],

$$\omega^2 = \left[ \omega_{LH}^2 + \frac{\bar{k}_\parallel^2}{1+\bar{k}_\perp^2} \Omega_e^2 \right] \frac{\bar{k}^2}{1+\bar{k}^2} \qquad (1)$$

where we define $k_x d_e = \bar{k}_\parallel$, $k_y d_e = \bar{k}_y$, $\bar{k}_\perp^2 = \bar{k}_y^2 + \bar{k}_z^2$, $\bar{k}^2 = \bar{k}_\perp^2 + \bar{k}_\parallel^2$, and $d_e = c/\omega_{pe}$ where $\omega_{pe}$ is the electron plasma frequency and c is the speed of light. The lower hybrid frequency is $\omega_{LH}^2 = \Omega_i \Omega_e / (1 + \Omega_e^2/\omega_{pe}^2)$ and in the dense plasma ($\omega_{pe} > \Omega_e$) of interest to the WD[1] simulation $\omega_{pe}^2 = 3\Omega_e^2$. In the limits: $\bar{k}_{\parallel}^2 > m/M$ and $\bar{k}_{\perp}^2 < 1$, Eq. (1) reduces to the dispersion relation of the regular electromagnetic whistler waves, $\bar{k}_{\parallel}^2 > m/M$ and $\bar{k}_{\perp}^2 >> 1$ it reduces to the electrostatic whistler waves (ESW), $\bar{k}_{\parallel}^2 / \bar{k}_{\perp}^2 < m/M$ and $\bar{k}_{\perp}^2 > 1$ it reduces to the electrostatic lower hybrid (LH) waves, and for $\bar{k}_{\parallel}^2 < m/M$ and $\bar{k}_{\perp}^2 < 1$ it reduces to the electromagnetic magnetosonic (MS) waves. In the 3D simulations[1] the box size was 120x24x24 in $d_e$ units. This implies that wave vectors can be generated in the simulation with a minimum of $\Delta \bar{k}_{\parallel} \sim 0.05$ and $\Delta k_{y,z} \sim 0.25$ and it is possible to initialize with $\bar{k}_{\parallel} = 0$ and $k_{y,z} = 0$.

Consider Fig. 14 of the WD simulation[1] that describes the 3D evolution. In the bottom left panel $|B_y^2|$ of the whistlers is localized in the region where $\bar{k}_{\parallel} \leq 0.1$, $0.25 \leq \bar{k}_y \leq 0.5$, which makes the angle of propagation to be $\theta \approx 75^o$. Thus, this band of whistlers propagate obliquely and not parallel to the ambient magnetic field $B_0 \equiv B_{0x}$. In the bottom right panel $|B_y^2|$ of whistlers are localized in the region where $\bar{k}_{\parallel} \leq 0.15$, $\bar{k}_y \leq 0.25$, which implies that this band of waves is a mixture of parallel propagating whistlers and whistlers with $\theta \approx 60^o$. Clearly, therefore, in the simulation mostly oblique waves are generated in the early phase. In oblique whistlers the electric field should be nearly along $\vec{k}$, while the magnetic field components should be in plane normal to $B_0$. Given that $E_y \approx E_z$ and $B_y \approx B_z$ all the time as seen in Fig. 5 of WD simulations[1], it may be inferred that turbulence of whistler as well as LH/MS waves with $B_x$ are isotropic in the plane normal to $B_0$ all the time. Based on this we see that the wave distribution at upper and bottom left panels in Fig.14 of WD[1] form a thin ring in the k-space ($k_\perp, \phi, k_\parallel$) where $k_y \equiv k_\perp$, centered around the brown spot with yellow halo.

From the upper left panel in Fig.14 in WD[1] at time (normalized by the ion plasma frequency, $\omega_{pi}$) near the saturation $t\omega_{pi} = 166$ it follows that the ion ring generates mostly oscillations with $\overline{k}_\parallel \leq 0.1$, $\overline{k}_y \approx 2.5$ and most intense waves are within an interval $\Delta \overline{k}_\perp \approx 0.25$. This is the signature of quasi-electrostatic LH/ESW wave turbulence. Using (1) we may calculate the frequency of these waves.

$$\frac{\omega^2}{\Omega_e^2} \approx \left[ \frac{m}{M(1+\Omega_e^2/\omega_{pe}^2)} + \frac{\overline{k}_\parallel^2}{\overline{k}_\perp^2} \right] \approx [0.75 + 0.8]10^{-3} \approx 1.5 \times 10^{-3} \quad (2)$$

$$\omega_{LH} = 0.027\Omega_e = \omega_{pi}/2, \ \omega = \omega_0 \approx 1.4\omega_{LH} \approx 0.04 \times 10^{-2}\Omega_e \quad (3)$$

At intermediate time $t\omega_{pi} = 221$ the upper right panel in Fig.14 of WD[1] indicates that the wave vectors have moved to the region where $\overline{k}_\parallel \leq 0.05$ and $1 < \overline{k}_y < 2$ making them LH/MS waves. These waves mainly have $E_\perp$ and $B_x \equiv B_\parallel$ components. But from the bottom left panel of Fig. 14 in WD[1] we see a magnetic field $B_\perp$ in the region where $\overline{k}_\parallel \leq 0.1$, $0.25 \leq \overline{k}_y \leq 0.5$ and during the same time (i.e., at $t\omega_{pi} = 221$) we see in Fig. 4 in WD[1] that $B_\parallel^2 \geq B_\perp^2$. This implies that both oblique whistlers and LH/MS waves are simultaneously present in this intermediate period of the simulation history.

It is important to note that despite $E_y = E_z$ as found in Fig. 5 in WD[1] simulation, the waves generated are not circularly polarized parallel propagating whistlers. This becomes obvious when we examine the energy partition in the electric, magnetic, and particle kinetic energies of the wave. It is well-known that in the LH/MS waves the ratio of the particle kinetic energy to the magnetic energy is $(1+\bar{k}_\perp^2)$ while the electron kinetic ($E \times B$ drift) energy plus electrostatic electric field energy is $(\omega_{pe}^2/\Omega_e^2)+1$, which is equal to 4 in the WD[1] parameters. Thus, we can conclude that the magnetic energy in the waves in Figs. 4 and 5 is much less than the sum of the electric field and particle kinetic energies all the time. For parallel propagating whistler the ratio of the wave magnetic energy to particle kinetic energy is $1/\bar{k}_\parallel^2 = \Omega_e/\omega > 1$. Thus, the electric field energy in Fig. 5 of WD[1] belongs mainly to the LH/ESW waves which were generated by the ring ions and are maintained isotropic in the plane perpendicular to the magnetic field by NL scattering by thermal particles as discussed in more detail below.

### 3. Nonlinear Analysis of the Simulation Results

In WD[1] it is suggested that waves in the top right and lower left panels in Fig. 14 is the result of nonlinear decay of the initially excited unstable LH/ESW waves with $\omega_0 \approx 0.04\Omega_e$ with $\bar{k}_\parallel \leq 0.1$ and $\bar{k}_y \approx 2.5$ (shown in the upper left panel) in two branches; one branch with $\omega \approx 0.02\Omega_e$ which is whistler waves in the region where $\bar{k}_\parallel \leq 0.1$, $0.25 < \bar{k}_y < 0.5$ (lower left panel); and another with $0.02\Omega_e < \omega < 0.035\Omega_e$ which is primarily LH/MS waves with $\bar{k}_\parallel \leq 0.05$, and $1 < \bar{k}_y < 2$ (the upper right panel). Subsequently, at time $t\omega_{pi} = 443$ whistlers with $\bar{k}_\parallel \leq 0.15$ and $\bar{k}_y \leq 0.25$ appear in the bottom right panel. WD[1] argues that these whistlers are the result of coalescence of two LH waves. This can only be partially correct because coalescence of two LH waves from the region where $\omega_0 \approx 0.04\Omega_e$, $\bar{k}_\parallel \leq 1$ and $\bar{k}_y \approx 2.5$ (the upper left panel), or where the frequency is in the range $0.02\Omega_e < \omega < 0.035\Omega_e$ with $\bar{k}_\parallel \leq 0.05$ and $1 < \bar{k}_y < 2$ (the upper right panel), will lead to daughter waves with frequency, $\omega_d$, which must be in the range $0.04 < \omega_d = 2\omega < 0.08\Omega_e$, since the daughter wave frequency must be the sum of the frequencies of the coalescing mother waves, each with frequency $\omega$. The frequency range of waves resulting from the coalescence should be larger than the prevailing frequency range of $\omega \leq 0.025\Omega_e$ for the parallel whistlers with $\bar{k}_\parallel \leq 0.15$ or oblique whistler with $\bar{k}_\parallel = 0.1$, $\bar{k}_y = 0.25$ found in the bottom right panel of Fig. 14 in WD[1]. Thus, the WD[1] interpretation of this part of the evolution as a wave-wave coalescence process is questionable because it violates the energy conservation law. However, coalescence is indeed possible for the waves from left bottom panel into parallel propagating whistlers in right bottom panel of Fig. 14 in WD[1] which covers the period of simulation history from $t\omega_{pi} \sim 221$ to 443.

A more interesting question is what is the nonlinear fate of the band of waves in the upper right panel with $0.02 < \omega < 0.035\Omega_e$, $\bar{k}_\parallel \leq 0.05$, and $1 < \bar{k}_y < 2$? This was not addressed in WD[1]. Examination of the nonlinear scattering rates using the simulation parameters indicate that these waves can scatter by thermal plasma particles. The nonlinear scattering rate for LH/MS/whistler waves, when $k_\perp^2 \gg k_\parallel^2$ and electrostatic contribution to the wave electric field $(\vec{E} \cdot \vec{k})$ dominates over the electromagnetic contribution $(\vec{E} \times \vec{k})$, for both wave-wave and wave-particle processes was estimated[2] to be,

$$\gamma_{NL} \sim \frac{\bar{k}_2^2}{1+\bar{k}_2^2} \sum_{k1} \frac{\omega_{pe}^2}{\omega_{k1}} \frac{|E_{k1}|^2}{8\pi n_0 T_e} \frac{(\vec{k}_1 \times \vec{k}_2)_\parallel^2}{k_{\perp 1}^2 k_{\perp 2}^2} \frac{\bar{k}_1^2}{(1+\bar{k}_1^2)}$$
$$\times \left\{ \left| \frac{\varepsilon_{k1-k2}^i (1+Z(\zeta_e))}{\partial(\varepsilon_{k1-k2}^i + \varepsilon_{k1-k2}^e)/\partial \omega_{k1}} \right| \frac{\gamma_{NL}}{\gamma_{NL}^2 + (\omega_{k1}-\omega_{k1-k2}-\omega_{k2})^2} + \mathrm{Im}\frac{\varepsilon_{k1-k2}^i (1+Z(\zeta_e))}{\varepsilon_{k1-k2}^i + \varepsilon_{k1-k2}^e} \right\} \quad (4)$$

$$\varepsilon_{k1-k2}^i = \frac{\omega_{pi}^2}{(\vec{k}_1-\vec{k}_2)^2 v_{ti}^2}\left(1+Z(\zeta_i)\right), \quad (5a)$$

$$\varepsilon_{k1-k2}^e = \frac{(\bar{\bar{k}}_1-\bar{\bar{k}}_2)^2}{1+(\bar{\bar{k}}_1-\bar{\bar{k}}_2)^2} \frac{\omega_{pe}^2}{(\vec{k}_1-\vec{k}_2)^2 v_{te}^2}\left(1+Z(\zeta_e)\right) + \frac{\omega_{pe}^2}{\Omega_e^2}\frac{1+(\bar{\bar{k}}_{\perp 1}-\bar{\bar{k}}_{\perp 2})^2}{(\bar{\bar{k}}_{\perp 1}-\bar{\bar{k}}_{\perp 2})^2}, \quad (5b)$$

$$\zeta_e = (\omega_{k1}-\omega_{k2})/(k_{1\parallel}-k_{2\parallel})v_{te}, \quad \zeta_i = (\omega_{k1}-\omega_{k2})/|\vec{k}_1-\vec{k}_2|v_{ti}, \quad (6)$$

where the subscripts '1' and '2' indicate mother and daughter waves and Z is the plasma dispersion function. As shown in Akimoto[3] et al., overlapping of cyclotron harmonics with $\gamma > \Omega_i$ renders the ions to be unmagnetized in the ion response, $\varepsilon_{k1-k2}^i$, and linear and nonlinear Landau resonance are possible. Hence these resonances and the consequent wave dissipation are expected in typical PIC simulations of the ion ring instability where the beam is instantly ionized as in WD[1]. (Note that in a realistic space experiment[4] where the ionization time could be tens of seconds (or larger) the instability growth rate will be much lower. For $\gamma < \Omega_i$ the ions will be magnetized. This makes ion Landau damping to be negligible and hence more energy can remain in the waves.) In particular, in the WD[1] simulation the ring instability growth is large $\gamma_{ring} \sim 0.1\omega_{LH} \sim \gamma_{NL,sat}^{k1\to k2} > \Omega_i$ and in the simulation the NL scattering of LH waves by thermal particles (ions and electrons) can be estimated as,

$$\gamma_{NL}^{k1 \to k2} \sim \frac{\omega_{pe}^2}{\omega_{k2}} \frac{\bar{k}_2^2}{1+\bar{k}_2^2} \sum_{k1} \frac{|E_{k1}|^2}{8\pi n_0 T} \frac{(\vec{k}_1 \times \vec{k}_2)_{\parallel}^2}{k_{\perp 1}^2 k_{\perp 2}^2} \frac{\zeta_i \operatorname{Im} Z(\zeta_i) + \zeta_e \operatorname{Im} Z(\zeta_e)}{4} \quad (7)$$

where equal ion and electron temperature is assumed, $T_e = T_i$. The maximum scattering rate correspond to the condition

$$\zeta_e = (\omega_{k1} - \omega_{k2})/(k_{1\parallel} - k_{2\parallel}) v_{te} \sim 1, \quad \zeta_i = (\omega_{k1} - \omega_{k2})/|\vec{k}_1 - \vec{k}_2| v_{ti} \sim 1. \quad (8)$$

In turbulent magnetosphere where the characteristic time of whistler spectra evolution is much longer than the ion cyclotron time there is no ion Landau damping and we have used NL scattering of LH waves by thermal electrons only[2,4]

$$\gamma_{NL}^{k1 \to k2} \sim \frac{\omega_{pe}^2}{\omega_{k2}} \frac{\bar{k}_2^2}{1+\bar{k}_2^2} \sum_{k1} \frac{|E_{k1}|^2}{8\pi n_0 T_e} \frac{(\vec{k}_1 \times \vec{k}_2)_{\parallel}^2}{k_{1\perp}^2 k_{2\perp}^2} \frac{\zeta_e \operatorname{Im} Z(\zeta_e)}{\left(1+(\omega_{k1}-\omega_{k2})^2/(\vec{k}_1-\vec{k}_2)^2 (T_e/M)\right)^2}. \quad (9)$$

In the WD simulation NL scattering by thermal electrons is not optimal because the simulation box is too small for the parallel wave length of the beat waves, $2\pi/(k_{1\parallel} - k_{2\parallel})$. However, NL scattering by thermal ions is optimal and can redistribute waves in broad $(\bar{k}_{\parallel}, \bar{k}_{\perp})$ volume along $\omega(k_{\parallel}, k_{\perp}) \approx const$, with only a small frequency decrease[2] and maintain an isotropic turbulence in the plane normal to $B_0$, as seen in the simulation[1] Figs. 4, 5 and mentioned earlier.

Now let us quantitatively consider the evolution of the spectra due to NL scattering of waves by thermal particles from the upper left panel in Fig. 14 of WD[1] where the cold ring generated waves dominate with $\omega_0 = k_0 v_r \approx 0.04 \Omega_e$ ($v_r$ is ring velocity) in the region around $\bar{k}_y \approx 2.5, \bar{k}_{\parallel} \leq 0.1$, with $\Delta \bar{k}_y \approx 0.25, \Delta \bar{k}_{\parallel} \approx 0.05$ to the upper right panel where $0.02\Omega_e < \omega < 0.035\Omega_e$, $\bar{k}_\| \leq 0.05$, $1 < \bar{k}_y < 2$ and to the lower left panel where

$0.015\Omega_e < \omega < 0.05\Omega_e$, $\bar{k}_\| \leq 0.01$, $0.25 \leq \bar{k}_y \leq 0.5$. WD[1] suggest the parametric decay[5] as the NL scattering process determining the nonlinear evolution of the waves generated by the ring ions. However, as pointed out earlier, the cold ion ring generates an ensemble of waves in a ring in k-space and not individual waves. Consequently, the parametric decay process[5], valid for individual "pump" waves, is not applicable to the WD[1] simulation. For wave-wave scattering in a weakly correlated ensemble of waves, as generated in the WD[1] simulation, we must consider the three wave coupling process[6] discussed below. We compare the rates of NL scattering by particles and the three wave NL coupling to determine the dominant process in the simulation. The unstable wave spectrum centered around $\omega_0 = k_0 v_r$ that is responsible for scattering by thermal ions should satisfy the resonance conditions (8) in which $|\vec{k}_1 - \vec{k}_2| \approx k_0$ and $(\omega_{k1} - \omega_{k2}) \approx k_0 v_{ti} \approx \omega_0 v_{ti} / v_r$ in the 3D case. Using this and the spectral width $\Delta\omega = (\partial \omega_{k1} / \partial \vec{k}_1) \Delta \vec{k}_1$ the unstable waves can scatter $\vec{k}_1 \to \vec{k}_2$ and the scattering rate by ions may be estimated as

$$\gamma_{NL}^{k1 \to k2} \sim 0.1 \omega_{LH} \frac{\bar{k}_2^2}{1+\bar{k}_2^2} \frac{\omega_{pe}^2}{\omega_{LH}^2} \frac{\langle E_{LH}^2 \rangle}{8\pi n_0 T_i} \frac{k_0}{\Delta k} \frac{\omega_0 v_{ti}/v_r}{\omega_0 v_{ti}/v_r + |\Delta \vec{k}_1 \partial \omega_{k1}/\partial \vec{k}_1|}, \qquad (10)$$

where the spectral density $|E_{k1}|^2$, which occupies the volume in k-space with $\Delta k_1 \ll k_0$ around $\vec{k}_0$, and the wave intensity $\langle E_{LH}^2 \rangle$ are related as

$$|E_{k1}|^2 \approx \langle E_{LH}^2 \rangle \frac{k_0}{\Delta k_1}. \qquad (11)$$

For parameters prevalent in the upper left panel of Fig. 14 in WD[1] simulation ($\omega_0 \approx 0.04\Omega_e$, $\bar{k}_\perp \approx 2.5, \bar{k}_\parallel \leq 0.1$, $\Delta\bar{k}_\perp \approx 0.25, \Delta\bar{k}_\parallel \approx 0.05$) we assume that the unstable spectrum evolves initially by mainly decreasing $\bar{k}_\parallel$ and $\omega_0 v_{ti}/v_r \ll |\Delta\vec{k}_1 \partial\omega_{k1}/\partial\vec{k}_1| \sim \omega_0$. This simplifies (10) as given in (17) below. Note that just a few scattering events can move the wave vector from the resonance ($\omega_0 = k_0 v_r$) with ring ions and place it into the stable region where $k_\perp < \omega_0/v_r$. Multiple such scatterings can move the waves into the LH/MS waves region ($\omega < \omega_{LH}, \bar{k}_2 \leq 2$), which is the condition found in the upper right panel of Fig. 14 in WD[1].

The decay rate of the unstable waves by three wave coupling process from the region around ($\omega_0, k_0$) into the LH/MS region ($\omega < \omega_{LH}, \bar{k}_2 \leq 2$) and into whistlers ($\omega_W = \omega_{k1} - \omega_{k2}, \vec{k}_W = \vec{k}_1 - \vec{k}_2$) is,

$$\gamma_{NL}^{\omega \rightleftarrows 2\omega} \sim \frac{\bar{k}_2^2}{1+\bar{k}_2^2} \sum_{k1} \frac{\omega_{LH}^2 \omega_{pe}^2}{\omega_{k1}(\omega_{k1}-\omega_{k2})} \frac{|E_{k1}|^2}{B_0^2} \frac{(\vec{k}_1 \times \vec{k}_2)_\parallel^2}{k_{\perp 1}^2 k_{\perp 2}^2} \frac{\bar{k}_1^2}{(1+\bar{k}_1^2)} \frac{\gamma_{NL}}{\gamma_{NL}^2 + (\omega_{k1}-\omega_{k2}-\omega_{k1-k2})^2}$$

$$\sim \frac{\bar{k}_2^2}{1+\bar{k}_2^2} \sum_{k1} \frac{\omega_{LH}^2 \omega_{pe}^2}{\omega_{k1}\omega_W} \frac{|E_{k1}|^2}{B_0^2} \frac{\bar{k}_{\perp W}^2}{\bar{k}_{\perp 2}^2} \pi\delta(\omega_{k1}-\omega_{k2}-\omega_W) \quad (12)$$

To obtain (12) we expanded $Z(\zeta_e), Z(\zeta_i)$ dispersion functions in Eqs. (5a,b) for large arguments $\zeta_e, \zeta_i$ and substituted in (4) so that,

$$\varepsilon^e_{k1-k2} + \varepsilon^i_{k1-k2} = -\frac{(\bar{k}_{z1}-\bar{k}_{z2})^2}{1+(\vec{\bar{k}}_1-\vec{\bar{k}}_2)^2} \frac{\omega_{pe}^2}{(\omega_{k1}-\omega_{k2})^2} + \frac{\omega_{pe}^2}{\Omega_e^2} \frac{1+(\vec{\bar{k}}_{\perp 1}-\vec{\bar{k}}_{\perp 2})^2}{(\vec{\bar{k}}_{\perp 1}-\vec{\bar{k}}_{\perp 2})^2} - \frac{\omega_{pi}^2}{(\omega_{k1}-\omega_{k2})^2} \quad (13)$$

and

$$\partial(\varepsilon^{e}_{k1-k2} + \varepsilon^{i}_{k1-k2})/\partial\omega_{k1} = \frac{\omega^2_{pe}}{\Omega^2_e} \frac{1+(\vec{\bar{k}}_{\perp1}-\vec{\bar{k}}_{\perp2})^2}{(\vec{\bar{k}}_{\perp1}-\vec{\bar{k}}_{\perp2})^2(\omega_{k1}-\omega_{k2})} . \qquad (14)$$

The rate of decay of unstable waves (from the upper left panel into LH/MS waves in the upper right panel and oblique whistler waves, $\omega_W = \bar{k}_{\perp W}\bar{k}_{\|W}\Omega_e$, in the lower left panel of Fig. 14 in WD[1]) given by Eq. (12) can be simplified to,

$$\gamma^{\omega \rightleftarrows 2\omega}_{NL} \sim \omega_{LH}\bar{k}^2_{\perp W} \frac{\omega^2_{pe}}{\omega^2_W} \frac{\langle E^2_{LH}\rangle}{B^2_0} \frac{k_0}{\Delta k_1} . \qquad (15)$$

The ratio of the rates of decay (15) and scattering (10) is,

$$\gamma^{\omega_0 \to \omega_w + \omega_{LH}}_{NL} / \gamma^{k0 \to k2}_{NL} \sim 10\beta \frac{\omega_0 v_{ti}/v_r}{\omega_0 v_{ti}/v_r + |\Delta\vec{k}_1 \partial\omega_{k1}/\partial\vec{k}_1|} \frac{\omega^2_{LH}}{\omega^2_W}\bar{k}^2_{\perp W} \sim \frac{10m}{\bar{k}^2_{\|W} M}\frac{\beta v_{ti}}{v_r} \sim \frac{\beta v_{ti}}{v_r} . \qquad (16)$$

Hence the NL scattering rate by particles in low beta plasma, as in the WD[1] simulation, is much larger than the decay rate.

Now we calculate the scattering rate (10) of LH waves into LH/MS waves on the upper right panel of Fig. 14 in WD[1] where $(0.7\omega_{LH} < \omega < 0.9\omega_{LH}, 1 \leq \bar{k}_2 \leq 2)$ using the saturated value of $\langle E^2_{LH}\rangle/8\pi nT_e = 0.015$ from the simulation provided in Table 1 in WD[1]. Here the frequency spectrum is relatively broad, i.e., $\Delta k \sim k_0$, which simplifies (10) to

$$\gamma^{k1 \to k2}_{NL} \sim 0.1\omega_{LH}\frac{\bar{k}^2_2}{1+\bar{k}^2_2}\frac{\omega^2_{pe}}{\omega^2_{LH}}\frac{\langle E^2_{LH}\rangle}{8\pi n_0 T_i}\frac{v_{ti}}{v_r} \qquad (17)$$

Using $\bar{k}_{\perp 2} = 1$, $v_r/v_{ti} = 7$ we get,

$$\gamma_{NL}^{LH \to LH/MS} \sim 0.2\omega_{LH}, \tag{18}$$

which matches reasonably with the saturation criterion $\gamma_{NL}^{LH \to LH/MS} \approx 2\gamma_{inst} \sim 0.1\omega_{LH}$ as discussed in Mithaiwala et al.[7]

We also note that the results of the 3D simulation[8] with $\beta = 0.03$, which is an order of magnitude larger than the WD[1] simulation, indicates that the saturated value of $\langle E_{LH}^2 \rangle / 8\pi nT$ remains nearly unchanged, implying that the saturated value of the waves are temperature dependent. The only difference in the parameters of the higher $\beta$ simulation[8] from the original WD[1] simulation was ten times larger temperature and accordingly three times larger ring velocity. The three wave NL coupling of LH-whistler waves in a low beta plasma, which does not depend on temperature as evident in Eq. (15), could not have maintained the ratio $\langle E_{LH}^2 \rangle / 8\pi nT$ nearly constant as found in the simulations with $\beta = 0.003$ and $\beta = 0.03$. Hence the only way that the ratio $\langle E_{LH}^2 \rangle / 8\pi nT$ remains unchanged is if the saturation mechanism is determined primarily by NL scattering by particles, as given in Eq. (10), which is a temperature dependent process.

## 4. Discussion and Conclusion

Quantitative analysis based on the simulation[1] parameters indicates that NL scattering by thermal particles contribute significantly to the evolution of the ion ring distribution. This was not discussed in the original WD article[1]. In addition to this there are other features in the WD simulation that are interesting and have not been highlighted. They are discussed below.

In the 3D PIC simulations[1] the spectral saturation occurs at wave energy density which is an order of magnitude smaller than in 2D as seen in Figs. 4 and 5 of WD[1]. This is because in 2D NL scattering by thermal plasma has the same rate for much larger wave energy[2]. And the ring relaxation rate in 3D is much less than in 2D as seen in Fig. 6 of WD[1] simulation because it is proportional to the saturated value of the wave energy. In addition, Fig.6a indicates that the ring continues to lose energy with nearly constant rate long after the fields have saturated implying that it is not quasi-linear diffusion that is responsible for saturation. These are consistent with the theory discussed in Mithaiwala et al.[7], which indicates that the relaxation of ring ions are affected by the nature of the nonlinear processes. It is, therefore, important to understand the dominant nonlinear process and include its effects in any predictive model of space plasma state.

In the WD simulation[1] the turbulent energy dissipates in ion and electron heating due to Landau damping. However in a realistic barium release experiment in the ionosphere[9] the neutral barium atoms ionize by solar ultraviolet radiation over the duration of its ionization time of 30 seconds creating an ion ring cloud a few hundred km in size. As a result the instability growth will be much lower than it is in the simulation[1] which employs instant ionization of all ring ions and hence release of all ring ion free energy instantly at time t=0 of the simulation. Since a realistic ring beam is spatially localized, the MS and/or whistler waves which appear due to NL wave decay and scattering by electrons will convect out of the region of localization of the ring cloud with roughly the Alfven speed in a tenth of a second which is much faster than the electron Landau damping time and hence avoid damping. Consequently, more energy will remain in the waves and this will result in the increase of the efficiency of energy conversion compared to that observed in the simulation. Again, this emphasizes the importance of realistic boundary conditions in simulations of actual space plasma phenomena.

Clearly, therefore, despite rapid advances in computing technology true simulation of space plasma phenomena continues to be a significant challenge. Indeed, as noted in WD[1], 3D simulation with extended-in-time ion release and wave convection out of the ring region is not yet practical. To circumvent such difficulties it is important to understand the dominant nonlinear issues of the problem of interest and parameterize them in tractable global simulations. An example of such a simulation model for the relaxation of an ion ring beam which is localized over a finite spatial extent was recently discussed by Scales et al.[10]. Nevertheless, the WD simulation[1] is very useful because it provides evidence of non-linear scattering processes in simulations which are the fundamental constituents of relaxation of a cold ring ion beam in the 3D space environment and hence extremely relevant for future experiments of VLF turbulence in space.

## Acknowledgments

We sincerely thank Dan Winske for numerous valuable discussions. Dan Winske and Bill Daughton have allowed us full access to their simulation data. We are very grateful to them for this and their continuous interest and encouragement of our work. This work is supported by the Naval Research Laboratory Base Program. One of the authors (L. Rudakov) acknowledges support from NSF grant AGS-1004270 at UCSD.